\documentclass[twocolumn]{revtex4}
\pdfoutput=1
\usepackage{latexsym}
\usepackage{amsmath}
\usepackage{times}
\usepackage{amssymb}
\usepackage{fancyheadings}
\usepackage[T1]{fontenc}
\usepackage[utf8]{inputenc}
\usepackage{url}
\usepackage{hyperref}
\usepackage{color}
\usepackage{graphicx}
 \usepackage{epsfig}

\begin{document}

\title{\bf The mechanisms of self-organised criticality  in 
  social processes of knowledge creation}
\author{Bosiljka Tadi\'c$^{a}$}
\author{Marija Mitrovi\'c Dankulov$^{b}$} 
\author{Roderick Melnik$^{c}$}

\affiliation{$^a$Department of Theoretical Physics, Jo\v zef Stefan Institute,
Jamova 39, Ljubljana, Slovenia }
\affiliation{$^b$Scientific Computing Laboratory, Center for the Study of Complex Systems, Institute of Physics Belgrade, University of Belgrade, Pregrevica 118, 11080 Belgrade, Serbia}
\affiliation{$^c$MS2Discovery Interdisciplinary Research Institute; M2NeT Laboratory and Department of Mathematics; Wilfrid Laurier University; Waterloo; ON; Canada\vspace*{3mm}}



\begin{abstract}
\noindent 
In the online social dynamics, a robust scaling behaviour appears as a key feature of many collaborative efforts that lead to the new social value. The underlying empirical data thus offer a unique opportunity to study the origin of self-organised criticality in social systems. In contrast to physical systems in the laboratory, various human attributes of the actors play an essential role in the process along with the contents (cognitive, emotional) of the communicated artefacts. As a prototypal example, we consider the social endeavour of knowledge creation via Questions\ \& Answers (Q\&A). Using a large empirical dataset from one of such Q\&A sites and theoretical modelling, we examine the temporal correlations at all scales and the role of cognitive contents to the avalanches of the knowledge creation process. 
Our analysis shows that the long-range correlations and the event clustering are primarily determined by the universal social dynamics, providing the external driving of the system by the arrival of new users.  While the involved cognitive contents (systematically annotated in the data and observed in the model) are crucial for a fine structure of the developing knowledge networks, they only affect the values of the scaling exponents and the geometry of large avalanches and shape the multifractal spectrum. Furthermore, we find that the level of the activity of the communities that share the knowledge correlates with the fluctuations of the innovation rate, implying that the increase of innovation may serve as the active principle of  self-organisation. To identify relevant parameters and unravel the  role of the network evolution underlying the process in the social system under consideration,  we   compare the social avalanches to the avalanche sequences occurring in the field-driven physical model of disordered solids, where the factors contributing to  the collective dynamics are better understood.
 
\end{abstract}

\maketitle
\thispagestyle{fancy}

\section{Introduction\label{sec-intro}}
In recent years, the self-organised criticality (SOC) is considered as one of the principal mechanisms responsible for the emergence of new features at a larger scale in various complex
systems.  The transition from the microscopic
interactions to the collective behaviour involves nonlinear dynamical phenomena when the system is driven out of equilibrium (for an overview of physical systems exhibiting SOC, see recent reviews in
\cite{SOC-reviewPhysRep2014,SOC-book2013,Avalanches-FunctMatGeophys2017} and the references there). 
In this context, SOC refers to the dynamical self-organization among the interacting units in response to repeatedly applied infinitesimal driving; the system's adaptation to the driving force leads to robust metastable states with system-wide correlations, fractal dynamics, and avalanches as the key signatures of criticality
\cite{SOC-book2013,Avalanches-FunctMatGeophys2017,BTW,dhar1990,tadicPRL1997}. 
 An avalanche is recognised as a mesoscopic dynamical structure consisting of a sequence of spatially connected events.  It has been newly pointed out that SOC plays a role in functioning of biological \cite{SOC-biology2011} and diverse other complex systems from neuronal dynamics
\cite{SOC-neuronalNat2007,SOC_neuroCulture2013} to animal behaviour \cite{SOC-birds2010} and  human history \cite{SOC-history2016}. 
 For instance,  the analysis of vast amount of the available brain
 imaging data and theoretical modelings provided the evidence that
 supports the SOC as an underlying mechanism of the brain functional
 stability \cite{SOC-neuronalNat2007,SOC-Brain-review2014}.

Although the avalanching behaviour and other signatures of criticality are readily observable in the empirical data of online social dynamics \cite{we-entropy,mitrovic2010a,mitrovic2011,we-MySpace11,we-emoRobotsCMP2014,virtual-world2017}, much less attention has been devoted to understanding the origin and the precise role of SOC in social systems. The key open question is
whether the social avalanches represent a unique class of
self-organised phenomena or, otherwise, they can be reduced to
standard models of physical SOC systems, describing the transition from the microscopic interactions to the observed complex spatiotemporal patterns.
Another interesting aspect of the problem concerns the interplay of the co-evolving network structure and the social dynamics that it supports. The question whether the SOC process shapes the structure, or the network
evolution enables the self-tuning towards the criticality
remains open.  Here, we address these issues by analysis of the empirical data of knowledge-creation social endeavours and using theoretical modelling.

In physics,  striking examples of the  multiscale dynamics characterising SOC are  observed in the turbulent flow
\cite{Aval-turbulence2001,SOC-turbulence2009,mfr-turbulence2013,SOC-book2013} and the kinetics of earthquakes \cite{corral2004,EQ-Nat2009}. The signatures of SOC
are also found in experiments with stressed granular materials
\cite{SOC-granular-stressed2002,StressGrannular-EQ2011},  driven disordered systems at hysteresis loop
\cite{berger_expBHN2000,koreans_films2011,martensites-SOC} and porous shape-memory alloys \cite{salje_2015,eduard_porous2014}. 
Furthermore, the avalanching dynamics is characteristic to the conduction in the assembled networks of living neurones in a solvent \cite{SOC-neuronsPRL2012} and nanoparticle films with single-electron tunnelings conduction \cite{we-chargetransport}, as well as  to the motion of topological
objects, such as vortices \cite{SOC-Vortex-PRB2004,SOC-Vortex-NatComm2014} and domain walls \cite{DWstochasticity_NatComm2010,DW-depinning-SciRep2014,BT_PhysA1999}.
The theoretical concepts were developed to describe the
emergence of collective behaviours from the microscopic interactions among many constitutive elements both in the classical and quantum systems \cite{book_StatPhys}. In this regard, often paradigmatic models were used as the instruments for investigations.  
Further, the use of scaling and renormalization group
ideas provided a better understanding of the role of
different scales in the dynamical systems driven away from the equilibrium \cite{UweT-NoneqRG2016,SOC-RG-Antonov2016,BT_disorderinducedcrit1998}.  
The precise description of the interactions in these physical systems allows investigating the microscopic mechanisms responsible for the triggering and propagation of an avalanche. The standard feature of all SOC systems is the accumulation of free energy, which then dissipates through the avalanches. While the energy source and triggering mechanisms are
physics-specific, common to all SOC systems is that the avalanching response is not in proportion to the forcing. Consequently, the power-law
distributions of the avalanches appear as one of the key features of SOC states. 
It has been recognised that the propagation of avalanches involves three phases \cite{SOC-book2013}: the initial growth
phase is supported by multiplicative chain reactions until the maximum dissipation is reached in the peak period, after which the activity is reduced and eventually diminishes, in the stopping phase. 
Often, the essence of SOC  can be captured by the elementary dynamics of sand-pile automata \cite{BTW,dhar1990}  on a two-dimensional lattice.
 In real systems, however, the presence of many physical parameters that can influence the dynamics makes it difficult to distinguish the potential SOC states (attractor
with a large basin of attraction) from the dynamical phase transition, which occurs by fine tuning of a relevant parameter.

In this work, we investigate the nature of  avalanches in the prototypal human collaborative endeavour of knowledge
creation \cite{we_SciRep2015}.  In this process, the knowledge and expertise of individual actors are transferred into a social value \cite{kimmerle,book_epistemology}--- the collective
knowledge, which is shared by all participants in the process.
In contrast to the physical systems in the laboratory, the human cooperation, as well as the new collective states, are evident \cite{we_SciRep2015,we-KCNets_PLOS2016}. Therefore, the empirical data on these social systems represent a valuable source to investigate the origin of self-organised criticality. On the other hand, certain attributes of
the human participants are crucial to the social cooperation; they remain elusive to the accurate theoretical modelling of the
interactions, which underlie the avalanche formation. In particular, the process of knowledge creation requires the appropriate expertise of the participating actors among other human attributes. Thus, the sub-dynamics representing the use of the communicated cognitive contents tends to constrain the social process itself.  In this regard, it remains unclear how these different aspects (social and cognitive) of the dynamics contribute to the appearance and propagation of the avalanches. 
To address this question, we combine the analysis of the empirical data from Questions \& Answers site Mathematics with the agent-directed modelling; the cognitive contents of each artefact are systematically encoded in the considered datasets. The agent's attributes are statistically similar to the users in the data, while their expertise is varied. The system is driven by the arrival of new agents. Our analysis reveals that the occurrence of avalanches is a robust social phenomenon, whereas
their fine structure, geometrical and fractal characteristics are
affected by the distribution of the expertise over the
actors. Furthermore, the interplay between the social and  knowledge processes is fuelled by the constant tendency towards the expansion of innovation. 
For a deeper understanding of the potential mechanisms, we use a comparison with the better controlled avalanching dynamics in physical systems. For this purpose, we analyse the model of a disordered system of the interacting spins, where the critical states at the hysteresis loop appear in the interplay between the  driving by the applied magnetic  field and the domain-wall pinning along the implanted magnetic (soft) and structural (hard) defects.  Although an analogy between spin alignments can be extended to knowledge matching among the social subjects, our objective here is different. We compare the fractal features of the avalanche sequences in both systems, which appear to be similar in a particular range of parameters of the physical system. These comparisons permit to identify certain factors of the social dynamics that are essential to the appearance of the collective state and can motivate further research towards a viable modelling of the social self-organisation.

In the following Section\ \ref{sec-KCP}, we introduce the essential characteristics of the processes of Q\&A and describe details of the agent-based model and the structure of the bipartite network that co-evolves with the social interaction.
Then the Section\ \ref{sec-avalanches} presents a detailed analysis of the knowledge-bearing avalanches both from the empirical data and simulations. The simulations and analysis of the avalanches in the driven spin system are given in Section\ \ref{sec-dds}. A summary of the results and the discussion are given in Section\ \ref{sec-discussion}.

\section{The stochastic process of knowledge creation and the
  co-evolving networks\label{sec-KCP}}
The knowledge creation via Questions\& Answers is a collaborative social endeavour, in which the knowledge of each participant is shared with others. By its nature, the interaction between these participants is indirect, mediated by questions and answers, in a way similar to user interactions via posted texts on Blogs \cite{mitrovic2010b,we_ABM_blognets,we-entropy}. Thus, the environment of the knowledge sharing can be represented as a co-evolving  \textit{bipartite network} with the \textit{actors} (users, agents)  as one partition, and the \textit{artefacts} (questions, answers), as the other partition \cite{we_SciRep2015}. In epistemology, to create a common value (knowledge) the \textit{meaningful social interactions} are required, in which the actor's response is adequate to
\begin{widetext}
\begin{figure*}[!htb]
\begin{tabular}{cc} 
\resizebox{36pc}{!}{\includegraphics{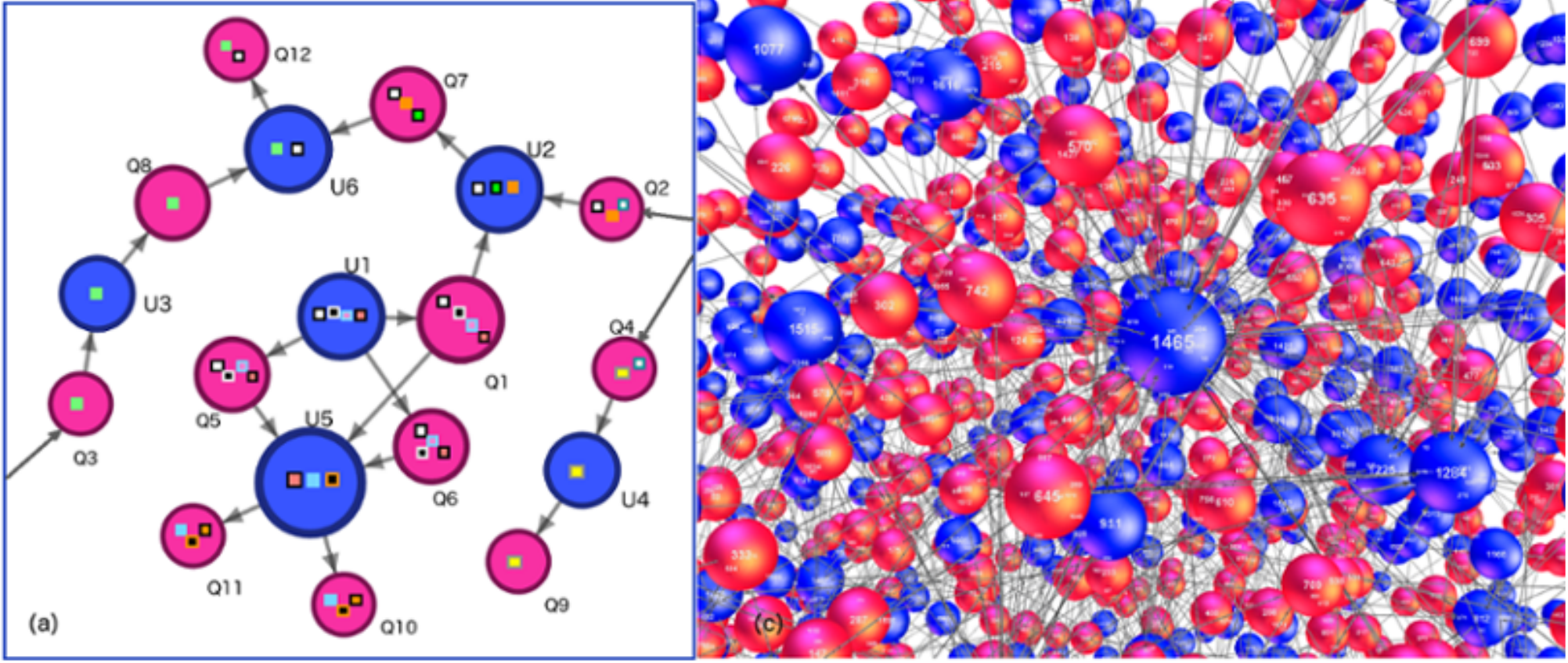}}\\
\resizebox{38pc}{!}{\includegraphics{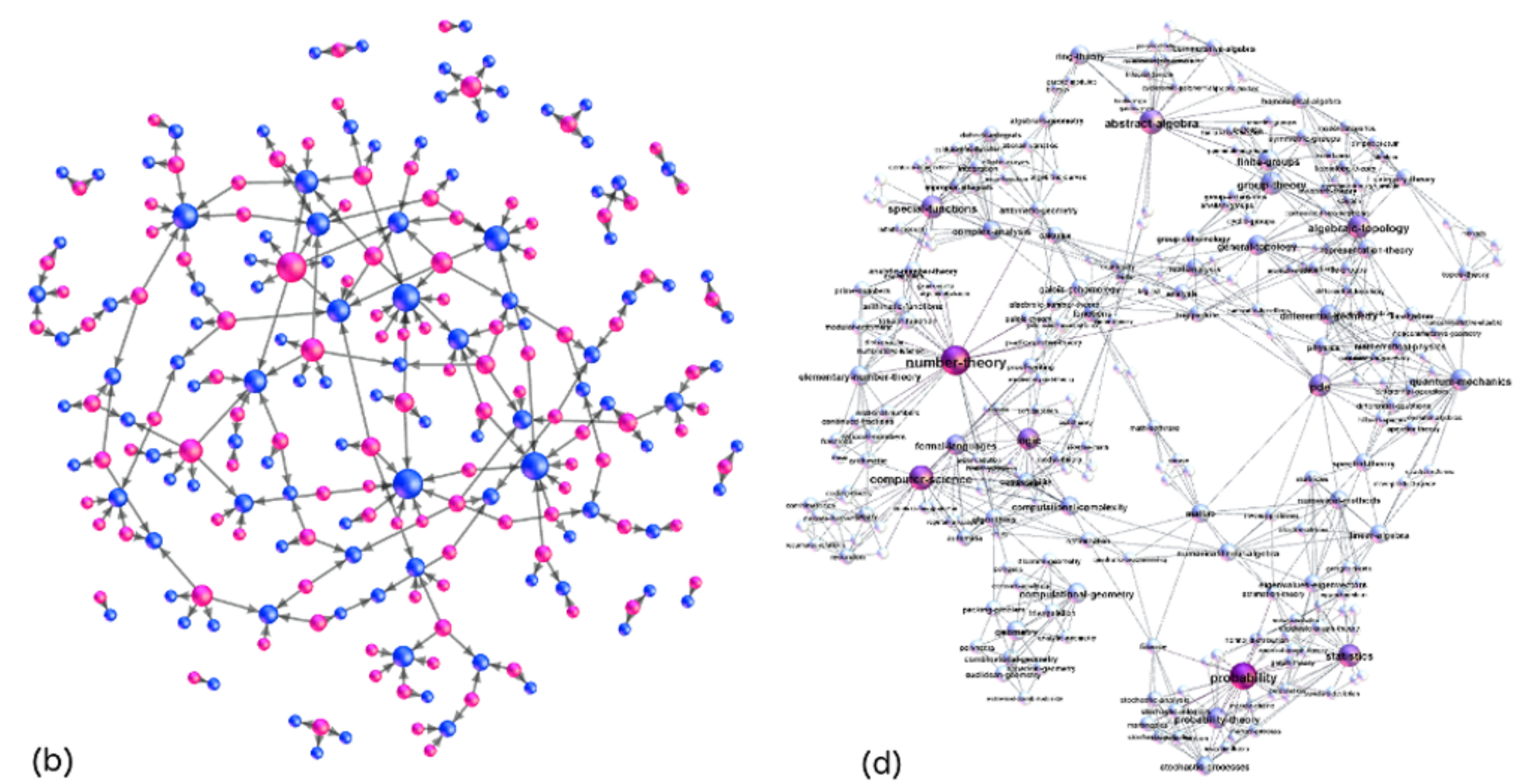}}\\
\end{tabular}
\caption{(a) Schematic view of the knowledge-exchange via Q\&A on a bipartite network;   boxes of different colours indicate the cognitive contents of the questions partition (red nodes) and the actor's expertise (in the blue node); a minimal matching between the expertise and the contents of the artefact occurs along the incoming links. Panels (b,c) show two bipartite networks extracted from the empirical data; the user nodes (blue) connect to question nodes (red) which compress all existing answers to that question, and the direction of the link indicates the question on which the user was active either by posting or answering it. Specifically:  (b) Innovation layer in the evolving bipartite network extracted at the end of year one, consists of the recently active questions and the users whose activity on  these questions occurred in the last  $T_0$=100 minutes and the nodes to which they were connected within the time depth of $6T_0$.  (c) A close up of the compressed bipartite network of users and the questions filtered such to contain the tag ``Linear Algebra'' among other tags; the network represents the  activity within the first two months of the considered empirical data. (d) The explicit-knowledge network containing the innovation tags attached to the tags of the year-one.}
\label{fig-KCP-schematic}
\end{figure*}
\end{widetext}
the needs of others \cite{kimmerle,book_epistemology}. Specifically,  the actor possessing an expertise can meaningfully act on the artefact where this particular expertise is required.  Thus, the essence of the dynamics is the contents-matching rule, as schematically illustrated in Fig.\ \ref{fig-KCP-schematic}a. 
A part of the evolving network extracted from the empirical data is also shown in Fig.\ \ref{fig-KCP-schematic}b.
Hence, the knowledge creation process consists of two mutually interconnected factors: the social dynamics and the constrained use of the cognitive contents.
The strict use of the available expertise in the knowledge creation processes is in marked contrast with the informal social communications on Blogs and similar systems, where the user's natural interests and emotions drive the activity \cite{we-entropy}.

\begin{figure}[!htb]
\begin{tabular}{cc} 
\resizebox{14pc}{!}{\includegraphics{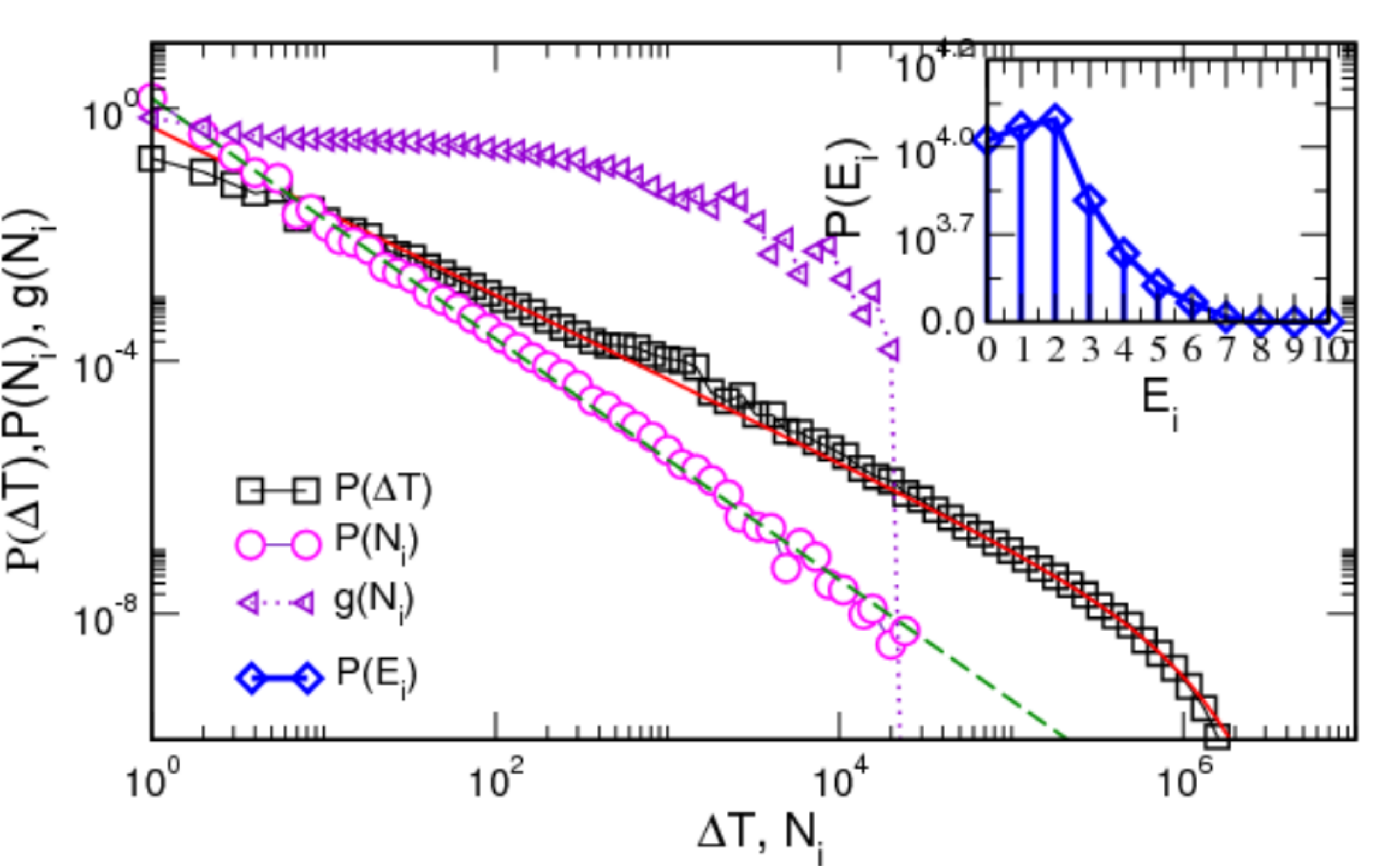}}\\
\end{tabular}
\caption{The distribution of the number of actions $P(N_i)$ per user, $g(N_i)$, and the delay time $P(\Delta T)$ of the users in the empirical dataset.  Inset: the distribution of entropy
  $P(E_i)$ based on the probability of the users' expertise; the data are extracted from the empirical set and used for designing the attributes of the agents.}
\label{fig-KCP-parameters}
\end{figure}

In the data from Q\&A site Mathematics, the cognitive content of each artefact is encoded by up to five tags, according to the standard mathematical classification scheme (MCS), for example, ``Graph Theory'', ``Probability'', ``Stochastic Processes'', ``Linear Algebra'', ``Algebraic Topology'', ``Differential Geometry'', and other. 
Whereas, the information about the user's expertise participating in the process can be inferred statistically, see the inset to Fig.\ \ref{fig-KCP-parameters}. This figure suggests a rather broad distribution of the expertise (i.e., different number of tags $2^{E_i}$) over the users in the native system. 
The main Fig.\ \ref{fig-KCP-parameters} shows another key characteristic of the experimental system: the user's heterogeneity in
the number of actions $N_i$ and a broad range of their interactivity times $\Delta T$. While the scaling exponents indicated by these histograms are system-specific, the prominent power-law decay of both quantities manifests the universally observed characteristics of the human behaviour online. 
Further patterns of the user's activity can be determined from the time stamp. 
Notably, a significant part of the actions of the present users is directed towards the issues posted by new arrivals. Occasionally, an active user looks for an older question with currently searched contents and brings it to the view of others.
The ratio of the posted versus answered questions $g(N_i)$ was found to depend on the number of actions of a user, cf. Fig.\ \ref{fig-KCP-parameters}. The very active users are in the minority; they often respond to the questions. On the other side, the majority of the least active users are engaged in posting questions and, by getting the satisfactory response, they disappear for a longer period.   The arrival of new users (referred to the
beginning of the dataset) is captured by the time series $p(t)$ in Fig.\ \ref{fig-Pt-randPt-example};  it
represents a stochastic process, which depends on the user's off-line life.   
Here, we adopt the interval of 10 minutes as a suitable time step for the pace of the activity in the system and use the empirical time series $p(t)$ to create the agents in the simulations. Then the reference time depth is $T_0=10$.

Observing the requirements for the minimal agent-based model (ABM) of the Web users \cite{BT-ABMbook}, we have introduced a model where the action rules and the attributes of the agents are taken from these empirical distributions, whereas their expertise can be varied \cite{we_SciRep2015}. 
Here, we briefly describe the main features of the model which is used for the simulations in this work. In particular, in each time step:
\begin{itemize}
\item \textit{agents are created}. The number $p(t)$ of new agents is created;  their profiles are defined by the number of actions $N_i\in P(N_i)$, the ratio of the posted vs. answered questions $g(N_i)$, and the expertise (according to the selected distribution); the new agents are placed on the active list; 

\item \textit{the agent's action performed}. Each agent from the active list either posts a new question or selects one from the list of recently considered artefacts to act on it.  The artefacts that are connected to the agent's network neighbourhood are looked first; with a fixed probability (0.5) the agent also finds a related question in the whole network, thus bringing it to the currently active context. In each case, the expertise matching rule applies; 

\item \textit{the active questions \& network are updated}. The list of active questions within time depth $T_0$  is maintained, and the network connections are updated according to the executed actions;  the agents linked to the questions on which the activity occurred within previous three steps are prompted for a new action;

\item \textit{new delay times are determined}. An agent gets a new delay time $\Delta t \in P(\Delta t)$ after every completed action or after being prompted for a new action;

\item \textit{the status of each agent is updated}. The number of actions of all agents is updated according to their activity, and the agents, whose number of steps reached the predefined $N_i$ are removed; the delay times of each remaining agent is updated, and each agent, whose delay since the previous event expires is placed on the active agents' list.
\end{itemize}
The expertise of an agent is a set of tags taken randomly from the list of 32 tags. The considered distributions of the expertise are $Exp1$ and $Exp3$, corresponding to a single-tag and three-tag expertise, and a broad range of the expertise $ExpS$, according to the empirical distribution shown in the inset in Fig.\ \ref{fig-KCP-parameters}.
For a comparison, we also consider a situation ($\mu$-process) where, instead of the actions described above in step 2, an agent finds an artefact in the entire system and acts on it with a fixed probability (0.25) while disregarding the expertise matching rule. Note that, in this case, the expertise matching can occur by chance; the agent's expertise is taken from the distribution $ExpS$.

 Considering the evolution of the system in \cite{we_SciRep2015}, we have shown that the process is characterised by the innovation growth with the number of events. In this context, the innovation is suitably defined as the number of new combinations of tags. The innovation is introduced into the system by new arrivals and the actions of the other agents through adding  their expertise to the accumulating contents on different artefacts. The innovation growth was observed both in the empirical data as well as in the simulations  \cite{we_SciRep2015}. It applies to various distributions of the expertise, excluding the case where all actors possess a strictly single-tag expertise. In this limiting case, the tag-matching rule prevents interlinking with other contents, thus leading to isolated communities that share the same tag.
In contrast, the innovation grows whenever at least a single actor possesses a combination of different contents in its expertise. The illustration in Fig.\ \ref{fig-KCP-schematic}a, for instance, indicates how the green tag of the artefact Q3 remains detached until the actor's U6 expertise combines it with the contents of Q7 and the actor U2 who posted it. The observed pace of the innovation growth depends on the distribution of the expertise (by the fixed activity patterns of the actors) \cite{we_SciRep2015}.
It is important to stress that, in the empirical data, the combination of tags in the expertise of each user already possesses a logical structure of mathematical knowledge. Hence, through these meaningful interactions that structure is preserved during the process. Consequently, the developing network of the used contents also exhibits the logical structure, as shown in \cite{we-KCNets_PLOS2016} by the community detection in the corresponding networks of tags;  an example of such an explicit knowledge network extracted from the same data is shown in Fig.\ \ref{fig-KCP-schematic}d.

\subsubsection{Growth of the bipartite networks by adding the  innovation
layers}
The interplay between the network structure and
the stochastic processes taking part on it consists the central problem in understanding the networks evolution and
their applications in various fields \cite{dorogovtsev-criticality,cherifi-networks,str-dyn2016}. 
Typically, the graph architecture represents geometrical constraints that shape the diffusion-like processes, likely to cause an anomalous diffusion. For example, the superdiffusion of the information packets \cite{TT2004} occurs on the correlated scale-free network when the traffic rules appropriately utilise the underlying structure. 
Some unusual situations arise when the structure evolves at the same pace as the SOC avalanching process on it. The random rewiring during the steps of the SOC dynamics has been shown to reduce the avalanche cut-offs \cite{savitskaya2016}, thus preventing a catastrophic event to occur. When the rewirings are strictly confined to the current avalanche area, the network appears to have the scale-free degree distribution, where the scaling exponent coincides with one of the avalanche size distribution  \cite{SOC-networkBTW2006}.
Other growth models that apply threshold-like constraints
inspired by SOC dynamics may yield the nonextensive features and scale-freeness  \cite{soares2005}.
In bipartite networks, however,  each partition plays a different role in the process, which leads to a more complicated structure--dynamics interplay. As mentioned above, these  types of networks often appear in the social dynamics on websites which maintain indirect communication between the users. The appropriate analysis of the artefacts mediating the users revealed \cite{we-entropy,mitrovic2010b,we_ABM_blognets,we_SciRep2015} how their emotional or cognitive contents affect the network from the node's degree to the mesoscopic community structure.

\begin{figure}[!htb]
\begin{tabular}{cc} 
\resizebox{14pc}{!}{\includegraphics{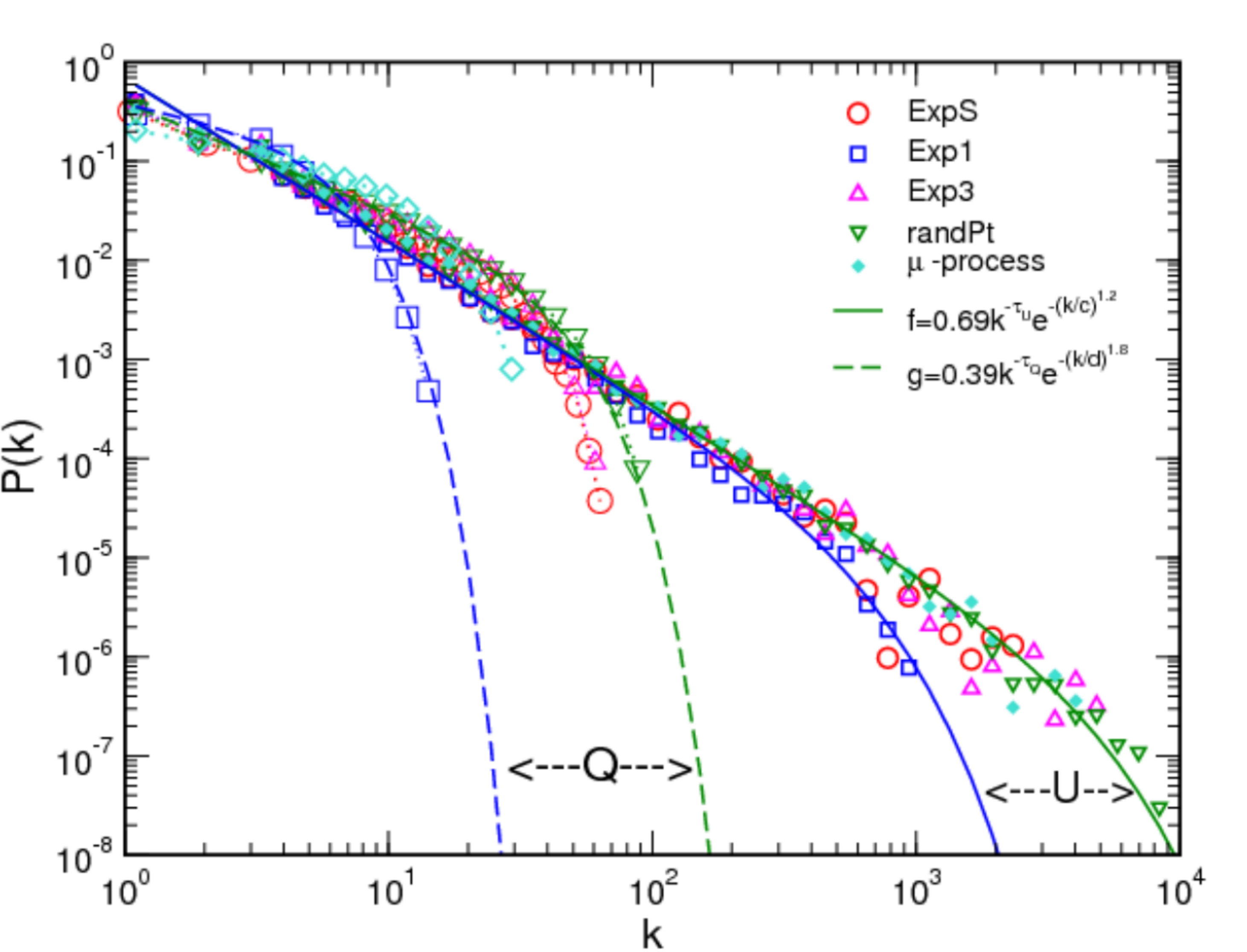}}\\
\end{tabular}
\caption{The degree distribution of the actors and question nodes in the bipartite networks for different expertise and driving indicated in the legend. The power-law fits with the exponent $\tau_U=1.58\pm  0.07$ and the stretching cutoffs in the range $c\in [508,3920]$ between two solid lines marked by $\leftarrow U\rightarrow$ are of the actor's degree. The distributions for the corresponding question nodes (indicated by the same type of symbol) have smaller cut-offs $d\in [5.8,42]$ between two dashed lines marked by $ \leftarrow Q\rightarrow$ and the exponent $\tau_Q=1.09\pm 0.1$.  In the case of $Exp1$ the exponents $\tau_Q\sim0.5\pm 0.16$ apply in a very narrow range, while  the exponential distribution fits the data for the $\mu$-process. }
\label{fig-Pq}
\end{figure}

In the Q\&A dataset that we consider here, the network growth as well as the pattern of activity of each user and the targeted questions can be extracted from the time stamp in the
data.
Moreover, to visualise the bipartite networks, we introduce a
compressed node which includes the question and all answers related to that question. A particular example of such compressed bipartite network from the empirical data is shown in Fig.\ \ref{fig-KCP-schematic}c. The corresponding networks from the simulated data exhibit the mesoscopic structure. The structure of communities sharing the emerging knowledge crucially depends on how the expertise is distributed over the involved participants, as it was shown in \cite{we_SciRep2015}.  Here, we are interested in the statistical properties of nodes in these bipartite networks. The results of the degree distributions are shown in Fig.\ \ref{fig-Pq}. They can be fitted by the power-law with  stretched-exponential cut-offs.
Statistically, the degree distribution of the agent's nodes in these networks follows the slope of the predefined number of actions $P(N_i)$, as expected, while their cutoffs depend on the expertise of the agents. In the case of the question-partition, the cases where the expertise-matching dominates exhibit a similar law but in a reduced range. In the meantime, the power-law behavior is reduced and  the cut-offs dominate  in the case $Exp1$ and $\mu$-process (see Fig.\ \ref{fig-Pq}). 

For the purpose of this work, it is interesting to recognise the \textit{innovation growth layer}, see Fig.\ \ref{fig-KCP-schematic}b, as the segment of the growing bipartite network where the most recent activity occurs.  As stated above, the new arrivals potentially bring the new combinations of the knowledge contents in their expertise,  which is expressed in the questions and answers. The set of currently active artefacts are posted or answered within a relatively small time window ($T_0$).  These nodes often occur at the outer layer of the network, see Fig.\
\ref{fig-KCP-schematic}b. Then the currently active users connect to these new artefacts while obeying the
expertise matching rule; thus, they connect the new contents to the issues of their previous activity and further through the network of the connected users and their artefacts in a given time depth. 
Therefore, the recently added artefacts connect the in-depth network via the active users and their previously established connections. Considering a particular time depth, for instance, 6$T_0$, with the  window $T_0$=100 minutes, we focus on the currently active layer of the network.  This layered growth of the bipartite network is fundamentally conditioned by the nature of  online human communications, where the latest posts appear on the top. Besides, at each event, an artefact older than the considered time depth is searched with a small probability and connected to the currently active matching contents.  The updated active layer then serves as an environment where the next arrivals often attach to, and so on. In the context of open dynamical systems, the addition of new users (or agents) and their artefacts can be seen as the \textit{driving mode} of this bipartite networked system.

\subsubsection{Extracting the time series and avalanches from the
  sequence of events}
\begin{figure}[!htb]
\begin{tabular}{cc} 
\resizebox{14pc}{!}{\includegraphics{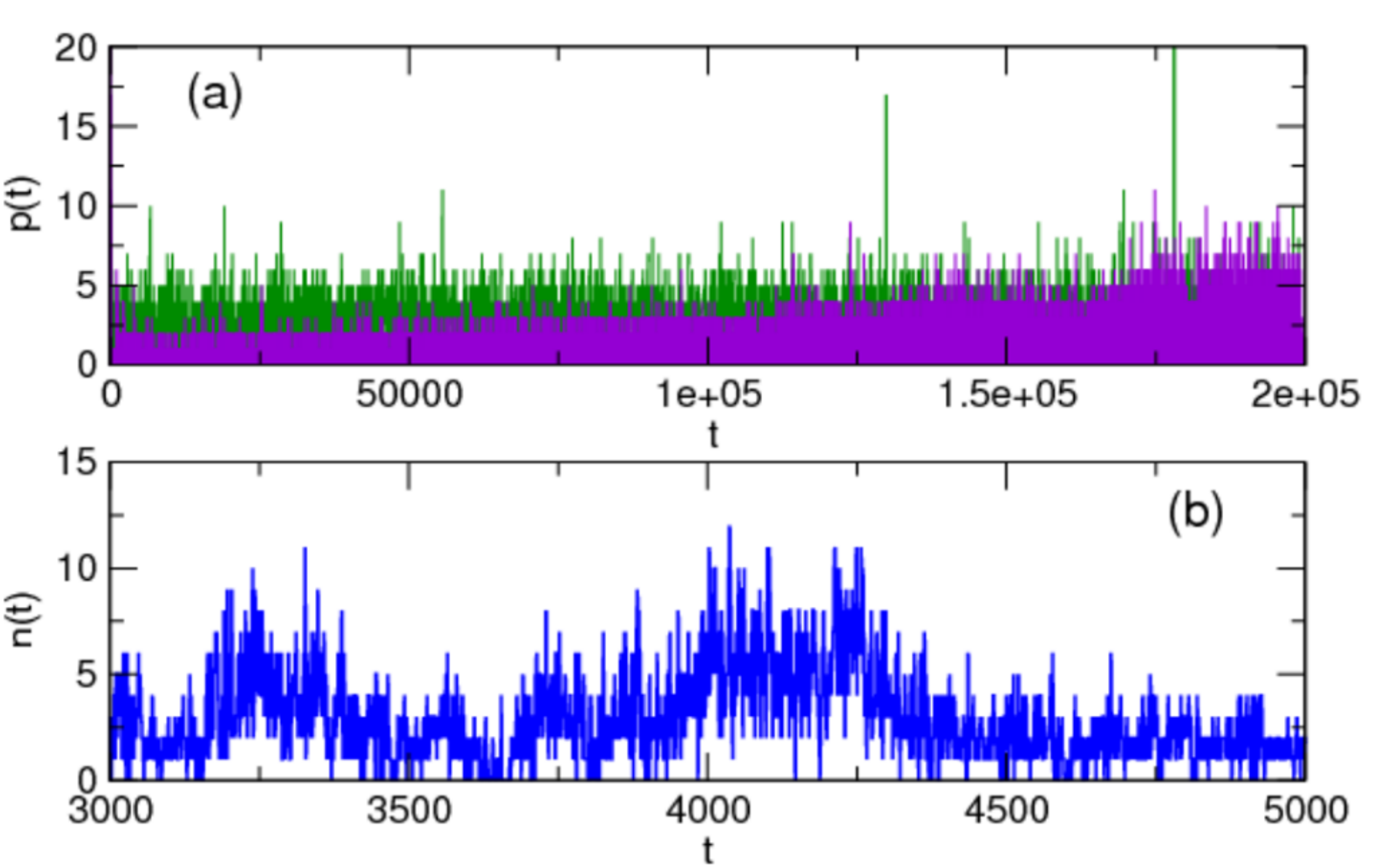}}\\
\end{tabular}
\caption{(a) The driving signal $p(t)$, front curve, and its randomized version, background curve.
(b) A part of the activity time series from the empirical data, exhibiting a number of large and small avalanches.}
\label{fig-Pt-randPt-example}
\end{figure}

Our focus in this work is on the avalanching behaviour which occurs as the network's response to the driving. Therefore, from the sequence of events in the empirical data or the simulated events, we first construct the corresponding time series that capture the fluctuations of the system's activity over time.  The possible occurrence of the clustering of events along these time series is a signature of the avalanching dynamics. For the illustration, an example of the time series with the avalanches is shown in Fig.\ \ref{fig-Pt-randPt-example}b. Specifically, an avalanche is identified as a
segment of the time series consisting of the data points $n(t)$ between two consecutive drops of the signal to the baseline (zero level). The two intersections of the signal with the baseline are recognised as the beginning $t_b$ and the end $t_e$ time of the avalanche. Then the avalanche size $S$ is given by the sum of all data points between the marked beginning and the end of the avalanche while the distance between these two points along the time axis defines the avalanche duration $T$, i.e.,
\begin{equation}
S=\sum _{t=t_b}^{t_e} n(t) \ ; T=t_e-t_b ; \
\label{eq-def-ST}
\end{equation}
As the example in Fig.\ \ref{fig-Pt-randPt-example} shows, the avalanches in the considered stochastic process of knowledge creation differ in size, duration and shape, closely reflecting the way that the activity propagates in the network. Moreover, a massive avalanche may follow immediately after a small one and vice versa. In the next section, we quantitatively study these features of the avalanches by considering the time series from both  the empirical data and the simulated events. 

\begin{figure}[!htb]
\begin{tabular}{cc} 
\resizebox{14pc}{!}{\includegraphics{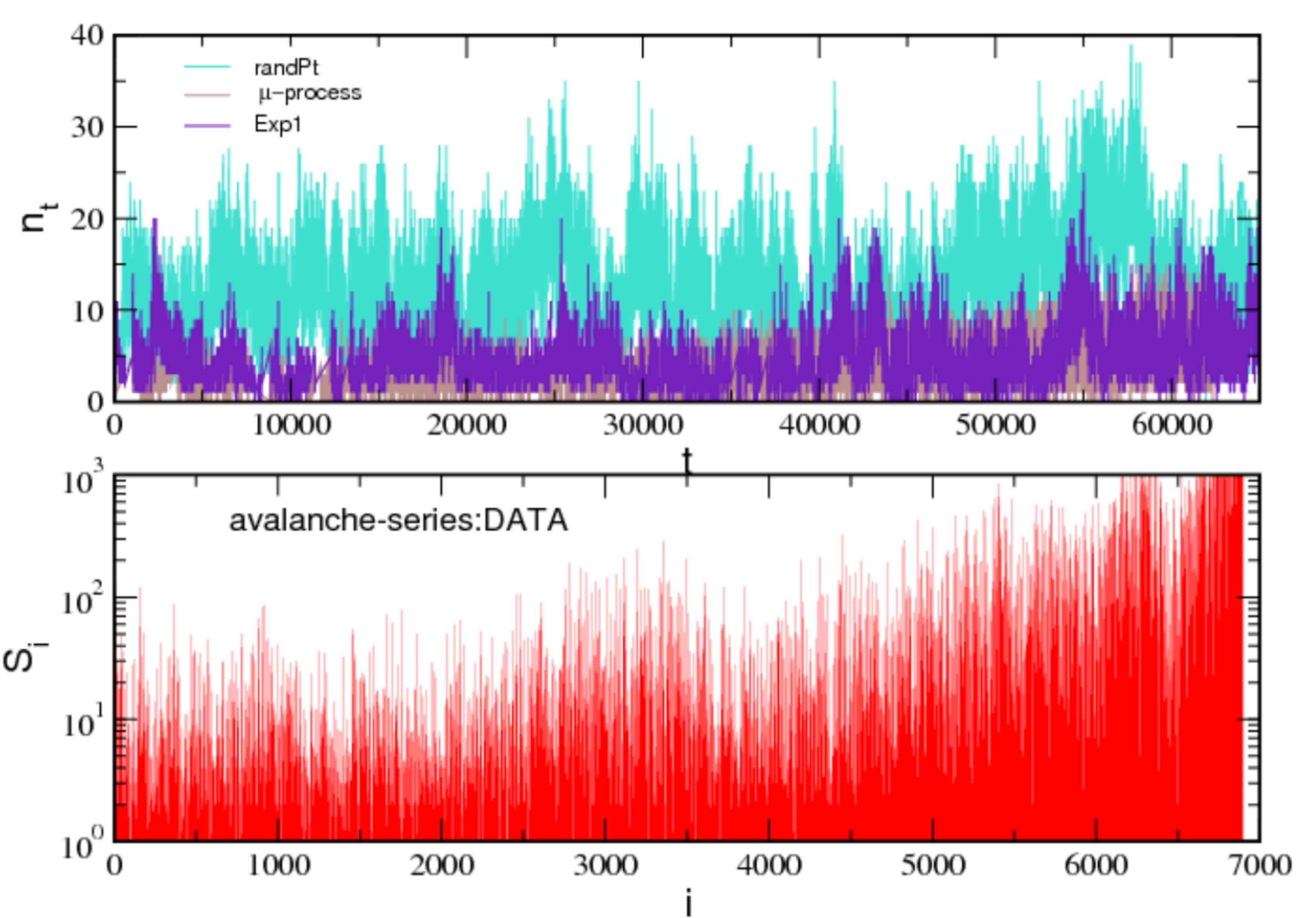}}\\
\end{tabular}
\caption{Top: Examples of time series of the activity simulated by agent-based model for the different expertise of the agents and $\mu$-processes. Bottom: Sequence of the avalanches determined in the empirical dataset.}
\label{fig-TS-avseries}
\end{figure}

Fig.\ \ref{fig-TS-avseries} top panel shows several examples of the time series of the number of event $n(t)$. Specifically, the time series indicated by $Exp1$ corresponds to the case when each agent possesses a single-tag expertise and these agents are added with the pace $p(t)$ as the new users appear in the real system.  The time series marked by $randPt$ is the system of agents with the distribution of the expertise $ExpS$ taken from the empirical data and driven by the randomised $p(t)$ signal.  Apart from a few high values at the start, the signal $p(t)$ exhibits an increasing trend, cf.\ Fig.\ \ref{fig-Pt-randPt-example}, which induces larger activity at later times. By randomising the signal, however, the larger values may occur randomly in time; thus, the number of added agents and their artefacts in the earlier times is larger than in the case of the original signal $p(t)$. This accelerated network growth mimics a larger driving rate in the context of SOC systems. The simulated data consist of 65536 steps, corresponding to the first 15 months of the real system time, where we find that 13045 users were active, posting 21998 questions and 179537 answers.
The bottom panel of the same figure shows the avalanche sequence determined from the empirical dataset.

\section{The structure of avalanches in knowledge creation
  processes\label{sec-avalanches}}
The use of fractal geometry and nonlinear analysis of time series has advanced the understanding of complex systems. Here, we employ the detrended multifractal analysis to study the time series of events as well as the sequences of the avalanches (clustered events) for various model parameters and the empirical data of the knowledge creation processes. 

\subsubsection{Temporal correlations and avalanche sizes\label{sec-temporal}}
\begin{figure}[!htb]
\begin{tabular}{cc} 
\resizebox{14pc}{!}{\includegraphics{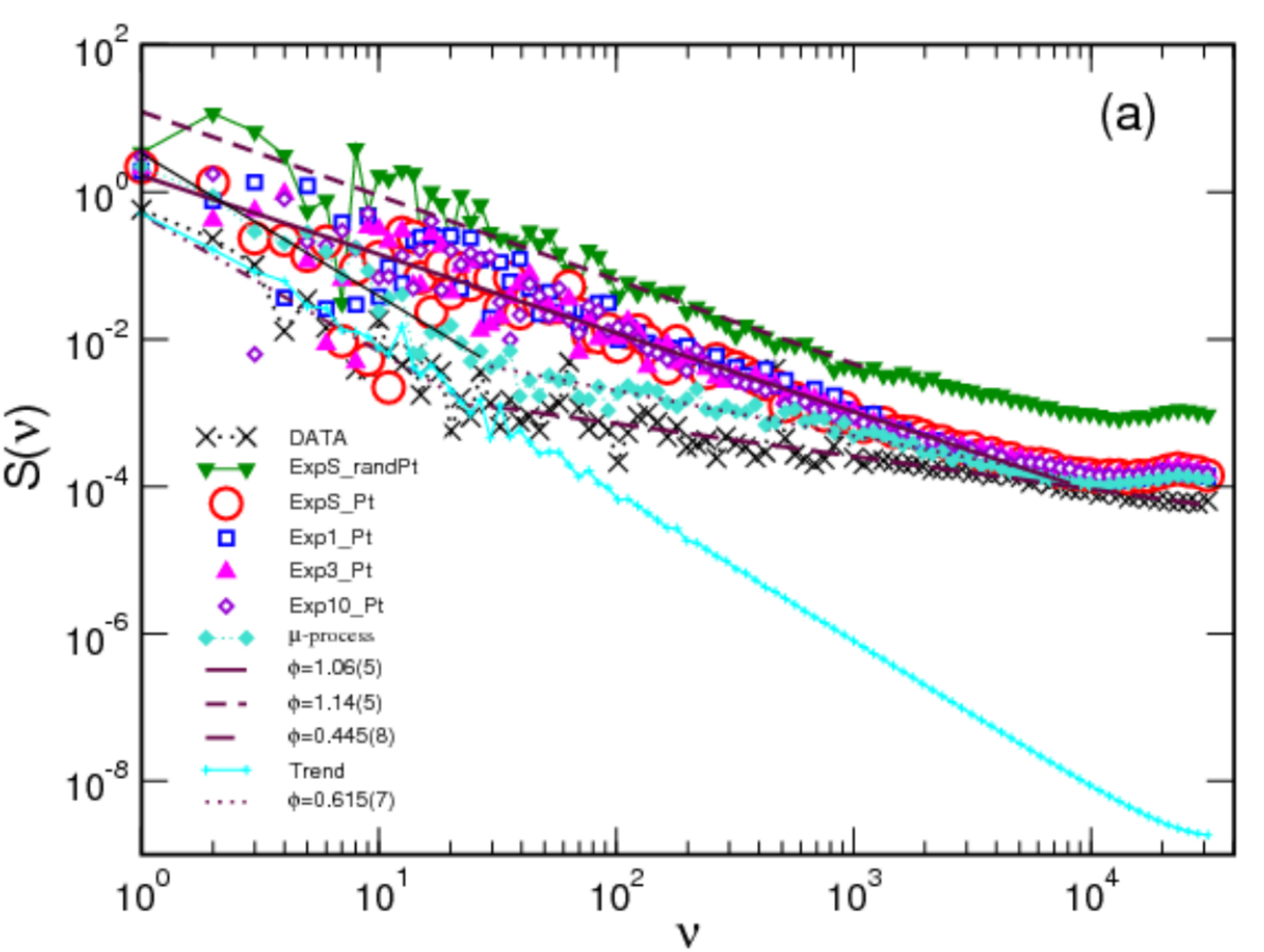}}\\
\resizebox{14pc}{!}{\includegraphics{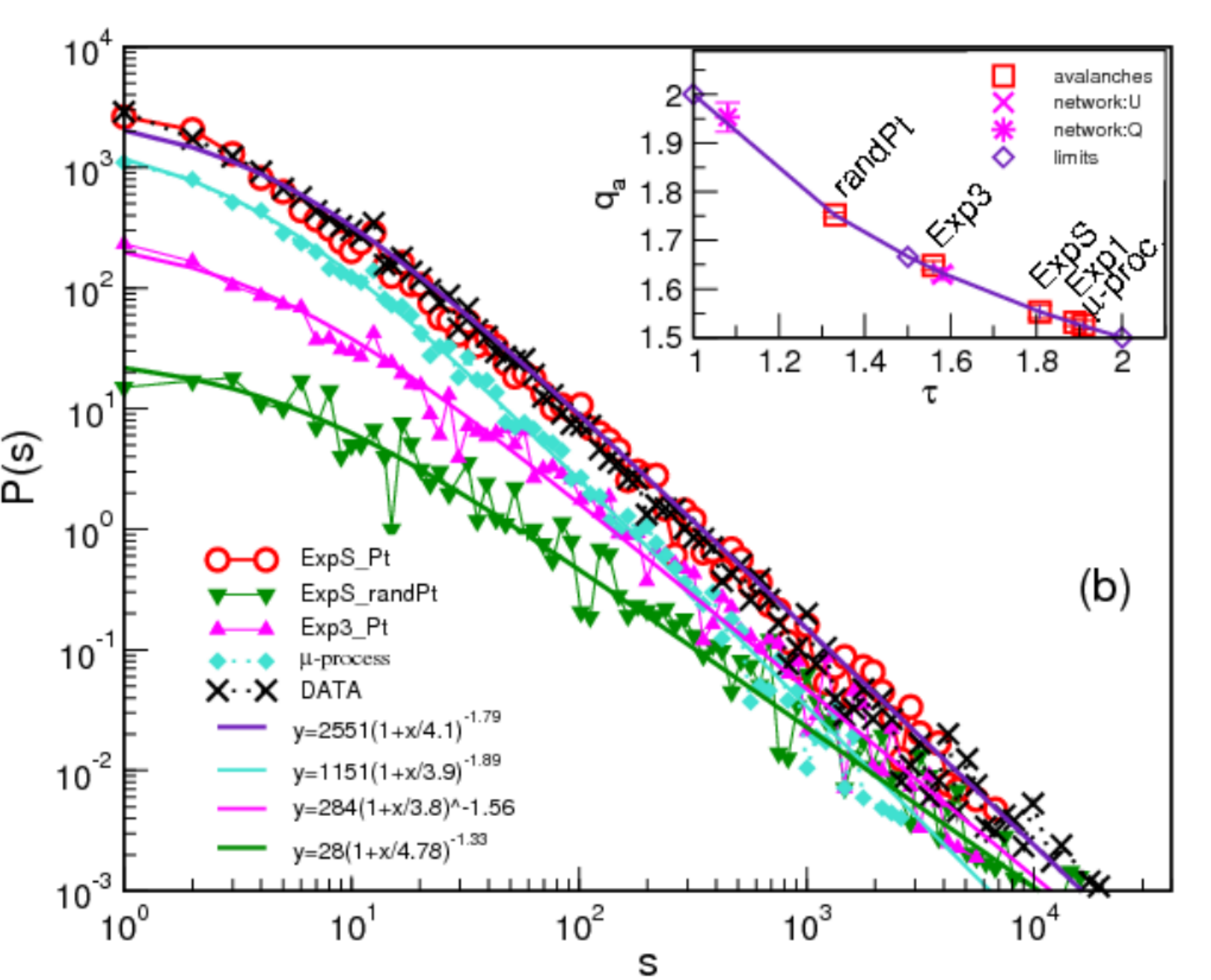}}\\
\end{tabular}
\caption{The power spectra (a) and avalanche size distributions (b)
in the empirical data and ABM for different expertise. Inset: The nonextensivity parameter $q_a$ vs $\tau_s$ for the considered cases.}
\label{fig-Size-Avalanches}
\end{figure}

The occurrence of avalanches in composite signals is not accidental but built on the temporal correlations at a larger scale. These correlations are manifested in the corresponding power spectrum as a power-law decay $W(\nu) \sim \nu^{-\phi}$, for a broad range of frequencies $\nu$. 
In Fig.\ \ref{fig-Size-Avalanches}a, we show the results for the power spectrum of the time series of events in the empirical data and the simulated signals for different agent's expertise and two driving modes.
The corresponding distributions of the avalanche sizes obtained from these time series are shown in Fig.\ \ref{fig-Size-Avalanches}b.
These figures indicate that an extended scaling range occurs over several orders of magnitude in the power spectrum as well as in the avalanche sizes. 
However, the driving mode and the actor's expertise affect the scale invariance in a different manner. Specifically, the increasing trend in the driving signal $p(t)$ is pronounced in the power spectrum of the empirical data and $\mu$-process.  Whereas, the increased activity in the innovation layer is balanced by the strict expertise-matching, resulting in the correlations of the flicker-noise type (middle curves). When the driving rate is elevated, i.e., $randPt$ case, the slope $\phi$ increases in the region of low frequencies and decreases in the high-frequency region (top curve).

While the shape of the driving signal is essential for the scaling in the power spectrum, the scale invariance of the avalanches is equally sensitive to the actual expertise of the agents.  The slopes of the distribution of the avalanche sizes $\tau_s$ are found in the range from 1.33 to 1.93, depending on the
expertise and driving; note that the degree distributions of user and question partitions in Fig.\ \ref{fig-Pq} are within this range (see also Discussion). However, the mathematical expressions that fit these distributions are different.  In particular, the distribution of avalanche sizes, shown in Fig.\ \ref{fig-Size-Avalanches}b can be fitted by the $q_a$-exponential function 
\begin{equation}
P(S)=A\left[1-(1-q_a)S/S_0\right]^{1/1-q_a} \ ,
\label{eq-q-exp}
\end{equation} 
where the parameter $q_a>1$ measures the degree of non-extensivity in the underlying stochastic process
\cite{tsallis1996,nonextensive-oxford,tsallis2011,pavlos_q-returns2014}. 
The values of the scaling exponents of the avalanche size distribution and the corresponding values of the non-extensivity parameter $q_a$ are shown in the inset in Fig.\ \ref{fig-Size-Avalanches}b. Notably, the distribution of the avalanche sizes obtained from the empirical data and simulations with the expertise of the agents $ExpS$ taken from the empirical distribution are similar and close to the case of
$\mu$-process. The probability of large avalanches increases with the increased average expertise, the case $Exp3$ is shown.  The accelerated network growth, such as in the case $randPt$,  induces an even bigger fraction of the large avalanches in comparison to small ones, consequently decreasing the slope of the distribution.

\subsubsection{Propagation and geometry of
  avalanches\label{sec-propagation}}
\begin{figure}[!htb]
\begin{tabular}{cc} 
\resizebox{18pc}{!}{\includegraphics{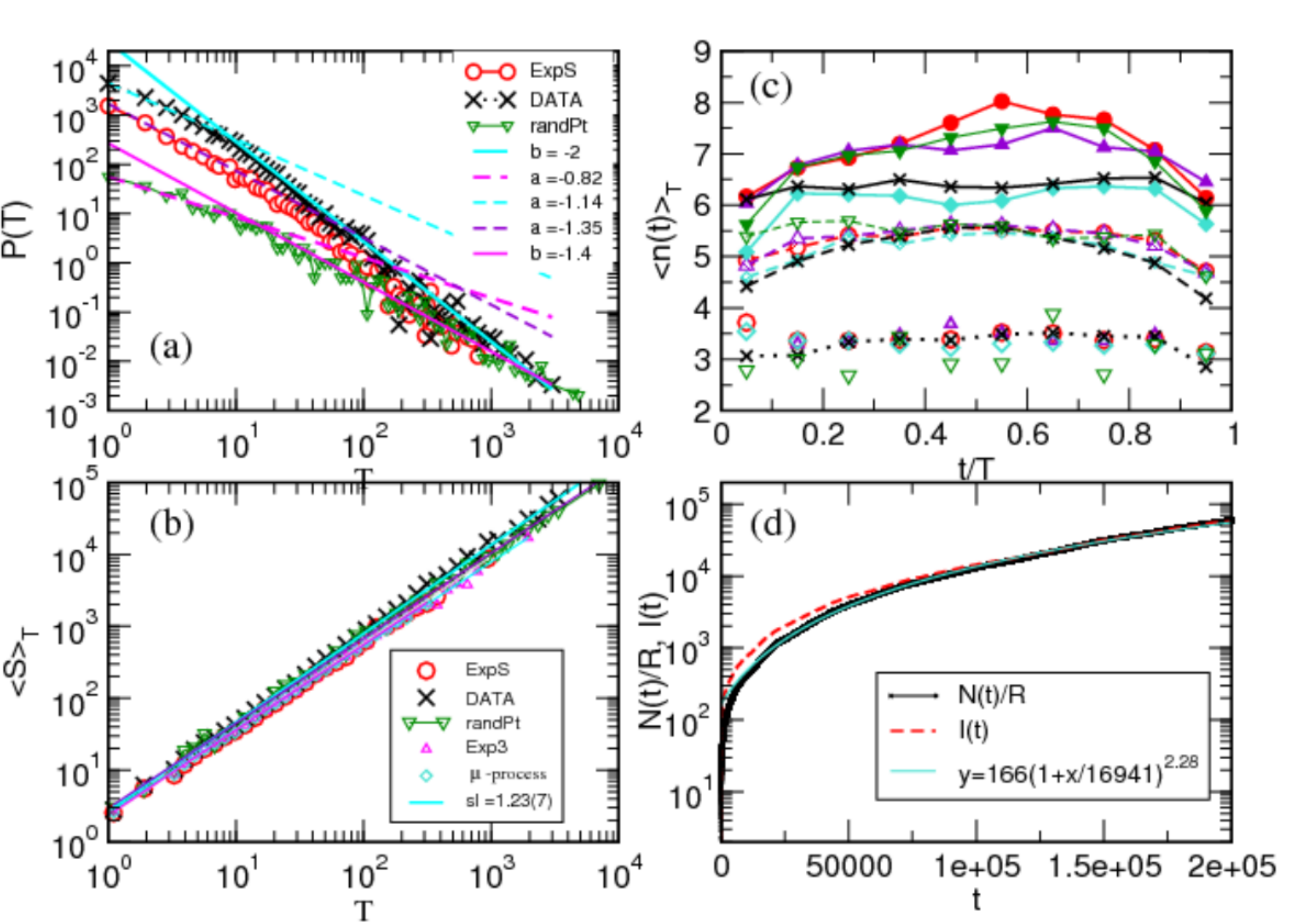}}\\
\end{tabular}
\caption{
(a) The distribution of avalanche duration and (b) the average
  size $\langle S\rangle _T$ of the avalanches of a fixed duration $T$ plotted against $T$ for
  the empirical data and simulations with varied expertise and driving
  rate, as indicated. 
 (c) The average
  activity $\langle n(t)\rangle _T$ within an avalanche of the
  duration $T$ averaged over the avalanches in three ranges: $T\leq 10$
  (half-filled symbols, dotted line),  $T\in(10,100]$ (open symbols, dashed lines), and
  $T>100$ (filled symbols, full lines). The symbol shape and color  correspond to the
  legend in the panel (b).
(d) Extracted from the empirical data, the innovation $I(t)$ increase with time and the integrated activity time series
$N(t)=\sum _t n(t)$, normalised by the ratio $R=\langle
n(t)\rangle/\langle dI(t)/dt\rangle$ =32.5.
}
\label{fig-AvTime-comb}
\end{figure}

Statistics of the avalanche sizes with power-law tails, as shown in Fig.\
\ref{fig-Size-Avalanches}b, is compatible with the occurrence of self-organized dynamics. In the following, we show that the propagation of avalanches and their shapes  further confirm these features of the underlying dynamics. The time-dependent characteristics of the avalanches evaluated for the above-studied sets of parameters are demonstrated in Fig.\ \ref{fig-AvTime-comb}(a-c).  Specifically, the distribution of the duration $T$ of avalanches, cf.\ Fig.\ \ref{fig-AvTime-comb}a, exhibits
different scaling behaviour for small avalanches with the duration $T<T_x\sim 10$ as compared to the asymptotic scaling law $P(T)\sim T^{b}$. In the asymptotic region, we find that (within the numerical error bars) $b=-2$ for the avalanche durations in the empirical data and a close value $b=-1.96$ for the simulated system
with $ExpS$.  Gradually smaller slopes are found for the cases $Exp3$ (not shown) and $randPt$. Similarly, the scaling exponent in the short-avalanche region vary with the expertise and the driving rate from $a=-0.82$, for $randPt$, to $a=-1.35$, for $ExpS$. The corresponding range for the empirical data is even shorter, see Fig.\ \ref{fig-AvTime-comb}a. Given different shapes, the size of the avalanches of a fixed duration can vary, cf.\ Fig\ \ref{fig-Pt-randPt-example}b  and Eq.\
(\ref{eq-def-ST}). Nevertheless, in the SOC systems, the average size$<S>_T$ of the avalanches of a given duration $T$ scales with $T$ as $<S>_T\equiv \sum_{t=t_b}^{t_b+T}n(t) \sim T^{\gamma_{ST}}$, where the exponent $\gamma_{ST}
=(b-1)/(\tau_s-1)$.
Fig.\ \ref{fig-AvTime-comb}b exhibits the plots $<S>_T$ against $T$ corresponding to the avalanches studied in this work. The average exponent $\gamma_{ST}
 =1.23\pm 0.07$ suggests rather narrow avalanches; apart from the duration range,  which varies with the simulation parameters, the variations of the exponent are rather small.

The precise shape of the avalanche of the duration $T$ is given by the sequence of the elementary pulses $n(t)$ over time. In Fig.\ \ref{fig-AvTime-comb}c, we show the average height $\langle n(t)\rangle_T$ belonging to the avalanche of a given duration $T$ evaluated in bins of the reduced time $t/T$. Three groups of avalanches are distinguished, in particular, the small avalanches of the durations $T\leq 10$, medium-duration $10<T\leq 100$, and large avalanches for the durations $T>100$. As the Fig.\ \ref{fig-AvTime-comb}c shows, the
shape of the small avalanches is practically independent of the system's parameters. The same conclusion applies for the medium-size avalanches in the peak region, whereas they slightly differ in the decaying phase and even more in the raising phase. In the case of the large avalanches, however, the major differences occur in the peak phase. Moreover, the peak shifts towards later times when the total expertise is increased, or a larger driving rate applied.

In the panel (d) of Fig.\ \ref{fig-AvTime-comb}, we show how the
total innovation $I(t)$ increases over time. The innovation, defined as the number of unique combinations of tags, is computed from the empirical data.
The total activity  (scaled by a constant factor) as a function of
time is also shown in Fig.\ \ref{fig-AvTime-comb}d. 
These results suggest that, apart from the excess of the innovation in the initial period,  the asymptotic law is the same for the total activity and the innovation growth with time.  While the average activity is 32.5 times larger, we can conclude that the activity is driven by the fluctuations of the innovation rate.  As mentioned earlier, the innovation is brought by the expertise of the actors.  Hence, the increase of the innovation can be considered as a driving force for the knowledge-creation processes.

\subsubsection{Avalanche sequences and multifractality\label{sec-MFR}}
\begin{figure}[!htb]
\begin{tabular}{cc} 
\resizebox{18pc}{!}{\includegraphics{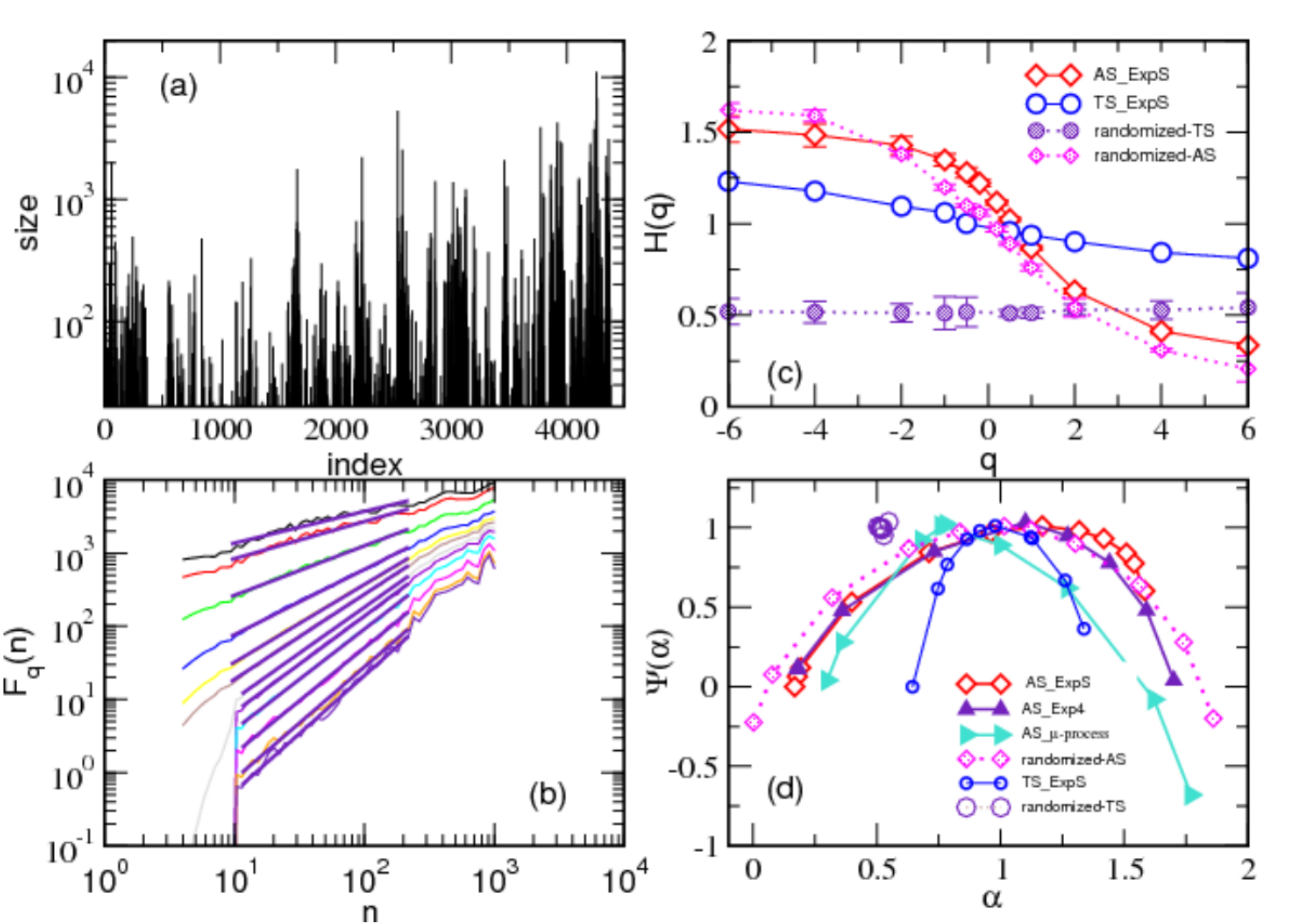}}\\
\end{tabular}
\caption{For the case $ExpS$, the avalanche sequence (a), the fluctuation function of the
  avalanche size series (b), and the generalised Hurst exponent $H(q)$ for the avalanche series and the underlying time series and their randomised versions (c). The panel (d) shows the singularity multifractal spectrum for the avalanche sequences determined for the cases with the varied expertise and driving, as indicated in the legend.}
\label{fig-Fluct-1}
\end{figure}

Compared to the time series of the number of events $n(t)$, the
avalanches are the objects occurring at a mesoscopic scale. Each avalanche consists of a certain number of the elementary pulses, cf.\ (\ref{eq-def-ST}); these pulses are combined in a different way to embody the growth,
peak and relaxation phase of the avalanche. 
Thus, the sequence of avalanches in time contains additional information about the nature of the underlying stochastic process. For instance, the avalanche return \cite{pavlos_q-returns2014,caruso2007} in many different complex systems reveals a non-Gaussian relaxation. For the knowledge-creation processes, we have demonstrated \cite{we_SciRep2015} that the avalanche
first-return statistics obeys $q_r$-Gaussian distribution with a large parameter $q_r\approx 2.45$.
 Here, we apply multifractal analysis to examine another signature of the complexity of these avalanche sequences.  In particular, we analyse the temporal sequences of the avalanche sizes $S_k$, where  $k=1,2,\cdots K_{max}$ is the index of the avalanche and $K_{max}$ stands for the total number of avalanches that occur in a particular time series. We
consider the avalanche sequences obtained for the combinations of the parameters studied in the preceding sections and the empirical data; two examples of such avalanche sequences are displayed in \ Fig.\ \ref{fig-TS-avseries} (bottom) and Fig.\ \ref{fig-Fluct-1}a.

To determine the multifractal spectrum $\Psi(\alpha)$ for the sequences $S_k$, we apply the detrended multifractal analysis (DMFA); we use the approach which was applied to different types of complex signals, as described in Refs.\  \cite{DMFRA2002,MFRA-uspekhi2007,dmfra-sunspot2006,BT-MFR2016}.
According to the standard procedure,  the profile of the signal is first constructed  by the integration 
\begin{equation}
Y(i) =\sum_{k=1}^i(S_k-\langle S\rangle)  \ .
\label{eq-profile} 
\end{equation}
The profile is then divided into $N_s=Int(K_{max}/n)$ non-overlapping segments of equal
length $n$. Then for each segment $\mu=1,2\cdots N_s$, the local trend $y_\mu(i)$ is
determined and the standard deviation around the local trend 
\begin{equation}
 F^2(\mu,n) = \frac{1}{n}\sum_{i=1}^n[Y((\mu-1)n+i)-y_\mu(i)]^2 
\label{eq-F2}
\end{equation}
is found.   
Similarly, the procedure is repeated starting from the end of the
signal, resulting in $F^2(\mu,n) =
\frac{1}{n}\sum_{i=1}^n[Y(N-(\mu-N_s)n+i)-y_\mu(i)]^2$ for $\mu
=N_s+1,\cdots 2N_s$. Combining the deviations at all segments, the
$q$-th order fluctuation function $F_q(n)$ is obtained  according to
\begin{equation}
F_q(n)=\left\{\frac{1}{2N_s}\sum_{\mu=1}^{2N_s} \left[F^2(\mu,n)\right]^{q/2}\right\}^{1/q} \sim n^{H(q)}  \ ,
\label{eq-FqHq}
\end{equation}
and plotted against the varied segment length $n\in[2,int(K_{max}/4)]$.
The scale invariance of $F_q(n)$ against the segment length $n$ is examined to determine the corresponding scaling exponent $H(q)$.  Here, the distortion parameter $q$ takes a range of real values. The main idea is that the segments of the signal with potentially different fractal features will be suitably enhanced by a particular $q$ value to become self-similar to the full signal and the corresponding scaling exponent $H(q)$ as a function of $q$ is measured.  Notably,  different \textit{small fluctuation segments} are
enhanced by the negative values of $q$, and the segments with \textit{large  fluctuations} dominate the fluctuation function  for the positive values of $q$.  In the limiting case of monofractal, $H(q)=H(q=2)$ is the standard Hurst exponent.
Using the scaling relation  $\tau (q)=qH(q)-1$, the exponent $\tau (q)$ of the box probability, defined in the partition function method \cite{DMFRA2002}, is computed.  Thus, the generalised Hurst exponents $H(q)$ can be related with
the singularity multifractal spectrum via Legendre transform  
$\Psi(\alpha)=q\alpha -\tau (q)$, where $\alpha =d\tau/dq$ is the singularity strength.

Fig.\ \ref{fig-Fluct-1}(a-c) shows the results for the case $ExpS$,
which incorporates the features of the empirical data and the expertise matching in the simulations. Specifically, the avalanche series, the corresponding fluctuation
function, and the generalised Hurst exponent are shown to demonstrate the procedure. Also, we show the results of the DMFA applied to the underlying time series for the same parameters. For the comparison, the analysis is performed for the randomised signals, the corresponding scaling exponents $H(q)$ are also depicted in Fig.\ \ref{fig-Fluct-1}c. 
Notably, both the time series of the activity and the related
avalanche series exhibit multifractal features. The span of the
generalised Hurst exponent is much larger in the avalanche
series. In the randomised case, the avalanche series exhibit almost unchanged multifractality while the time series of pulses becomes a monofractal with the properties of white noise ($H(q)=0.5$ within the numerical error bars). These findings indicate that the origin of the multifractality in the time series may be found in the temporal correlations,
which are of the $1/\nu$-type, see Fig.\ \ref{fig-Size-Avalanches}a.
While, for the avalanche series, the scale-invariance of the
distribution of sizes can be the sole reason for the
observed multifractality. In this regard, it is interesting to
analyse the width of the singularity spectrum $\Psi (\alpha)$ for different parameters of the model. 
These results are shown in Fig.\ \ref{fig-Fluct-1}d. In agreement with the values of $H(q)$ for the case $ExpS$ in the panel
(c), the spectra corresponding to the avalanche series and the randomised avalanche series are wide. In contrast,  the multifractality of original time series results in the narrow range; further, the spectrum is reduced to a close vicinity of the point $\Psi(\alpha=0.5)=1$, corresponding to the monofractal randomised time series. 
For the varied expertise of the actors, we obtain wide spectra of the avalanche series, where the singularity strength $\alpha \in [0.2,1.7]$, cf.\ Fig.\ \ref{fig-Fluct-1}d. Notably, the various parameters of the model mostly affect the right side of the spectrum, corresponding to $q<0$ region, i.e., small fluctuation segments in the avalanche series. While, the variations are less pronounced for the
large fluctuation segments, appearing in the left end of the spectrum $\Psi (\alpha )$. In this end, the
empirical data and model simulations lead to similar results. These findings indicate how the sequences of small and large avalanches are affected by the available expertise; the occurrence of the large fluctuations in the avalanche sizes might be chiefly conditioned by the number of the actors involved.

\section{The impact of vacancies on multifractal spectrum: A comparison to  spin-alignment avalanches\label{sec-dds}}
As mentioned in the Introduction, the complexity of the social
interactions prevents the exact description of the mechanisms at the elementary scale, calling for a comparison to better-understood physics models. 
In this regard, the interacting spin system described by the
random-field Ising model at zero temperature and
driven along the hysteresis loop represents a paradigm of complex dynamical behaviour far from the equilibrium \cite{reche-originPRE2013}. 
In this example, the spin alignment along the slowly increasing
external field is balanced by spin-spin interactions and the local
constraints due to the random-filed disorder. 
The dynamics of spin flips under the disorder induced constraints  and interactions was often employed to model opinion formation \cite{sociodynamics-book2006}, processes driven by social-balance \cite{SOC-socialBalance2011} and other cases.
For the purpose of this work, we aim to explore the impact of vacancies in the underlying network onto the multifractal spectrum of the avalanche sequences.  We consider spin-alignment avalanches in the zero-temperature
random-field Ising model ZTRFIMc with two-state spin
site $S_i(t)=\pm 1$ at each lattice site $i=1,2,\cdots N$ with a fraction $c>0$ of defect sites where the spin is absent. 
The energy ${\cal{H}} = -\sum _i\tilde{h}_i(t)S_i(t)$ is minimised by
the spin alignement along the current value of the local field
$\tilde{h}_i(t)$, where
\begin{equation}
\tilde{h}_i(t)=\sum_{j\in nn}J_{ij}S_j(t) +h_i +B(t)\ ;  S_i(t+1) =
sign \tilde{h}_i(t)
\label{eq-si}
\end{equation}
Here, $h_i$ is the local quenched random
filed, which is described by Gaussian distribution of zero mean and
the width $\Delta$. The system is driven by the slowly increasing external field $B(t+1)=B(t) + \delta B$ starting from a large negative value.
The ferromagnetic interaction $J_{ij}$ among the pair of
spins at the adjacent sites $i,j$ has the positive mean $<J_{ij}>=J$ and the second cumulant $Jc(1-c)$, where  $c>0$ is the probability that the spin is absent at a randomly selected site.  The dimensionless parameters $f\equiv \Delta /J$ and 
$r\equiv \delta B/J$ characterise the pinning strength and the driving rate, respectively.

Given the domain structure in these disordered systems, the magnetisation reversal occurs in a series of jumps by the slow field ramping along the hysteresis loop. These magnetisation changes are directly related to the motion of the domain walls, accompanying the expansion of the domains which are oriented parallel to the field. The size of the magnetisation changes thus occur in the interplay between the driving by the external field and pinning of the domain walls at by the local random fields, oriented opposite. These magnetisation changes in time represent the data points in the  Barkhausen noise,  a complex time series from which then the avalanches can be determined. The scale-invariant behaviour of the
Barkhausen avalanches and their dependence on the strength of the random-field disorder has been well understood \cite{eduard_FSS2003,eduard_spanning2004,reche-originPRE2013,djole_spanningaval2014,reche-originPRB2016}. Recently, it has been shown \cite{BT-MFR2016} that the Barkhausen
noise exhibits multifractal structure. Moreover, the dynamical regime in the central part of the hysteresis loop, where large avalanches can occur for the weak disorder, has a significantly different spectrum from the dynamics at the beginning of the hysteresis loop. In contrast, the presence of hard defects in ZTRFIMc has been much less investigated. 
 Specifically, even in the weak random-field pinning that allows system-wide avalanches, the presence of hard defects induces a characteristic length, which affects the cutoff size of the avalanches and also the universality of the scaling exponents \cite{tadic1996,BT_EPJB2002}. Here, we are interested in the
 dynamics of the avalanches in the presence of site defects.

\begin{figure}[!htb]
\begin{tabular}{cc} 
\resizebox{18pc}{!}{\includegraphics{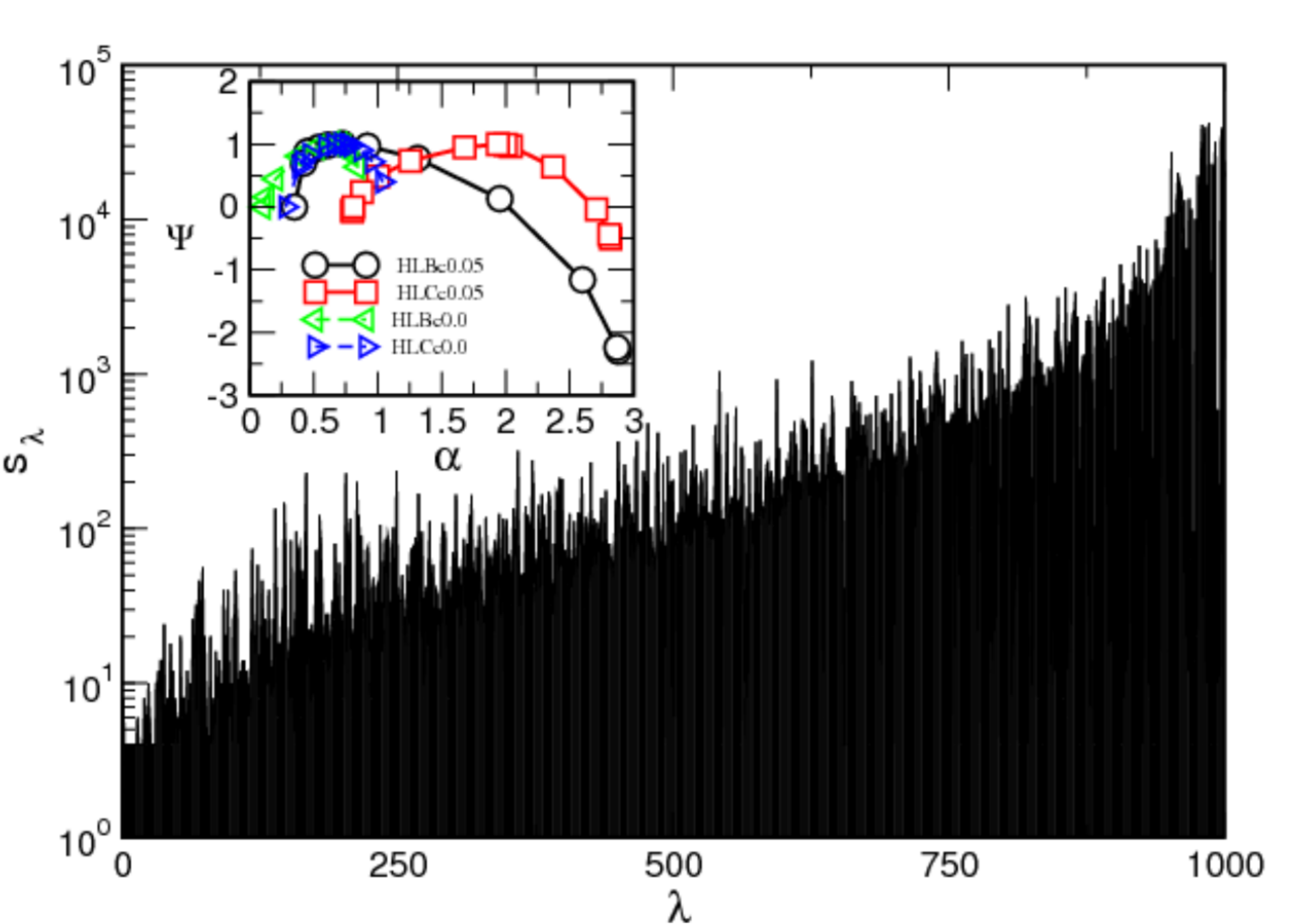}}\\
\end{tabular}
\caption{The avalanche sequence at the beginning of the hysteresis loop
  (HLB)  of the ZTRFIMc for $c=0.05$ and
  weak random field pinning; Inset: the singularity multifractal spectra in the
  hysteresis loop center (HLC) and the loop beginning (HLB)  for $c=0.05$
  and $c=0.0$. }
\label{fig-bhnavseries-MFR}
\end{figure}

We  consider a small concentration of the randomly dispersed site defects $c=0.05$  on top of the weak random field disorder and slow driving; thus we use a representative set of parameters in this regime \cite{BT-MFR2016}:
$f=2.3<f_c$ and $r=0.02$ in the 3-dimensional cubic lattice of $100^3$ spins. In the absence of the site defects ($c=0$), the system of this dimension would undergo a domain-wall depinning via large avalanches in the central part of the hysteresis loop \cite{BT-MFR2016}. However, the small percentage of site defects suffices to perturb this critical behaviour by the pinning of the domain walls at a distance $\ell \sim 1/c$; whereas,  at the distances $x\lesssim \ell$ the domain wall motion is accelerated by the external field, corresponding to the regime of the weak random-field pinning. 
Consequently, the small avalanches are similar as in the case of weak pinning without site defects, while the propagation of large avalanches is considerably hindered. 
These effects are reflected in the multifractal spectrum of the avalanche sequences, in particular, by increasing the difference in the generalised Hurst exponent for $q>0$ and $q<0$ values. The resulting singularity multifractal spectra are shown in the inset to Fig.\ \ref{fig-bhnavseries-MFR}. Although the avalanches tend to be larger in the central part of the hysteresis loop, the presence of site defects induces a mixture of small and large events, which results in the substantial broadening of the spectrum in comparison to the case of $c=0$.

It is interesting to point the similarity between the spectra in the avalanche series that are affected by site defects in 
Fig.\ \ref{fig-bhnavseries-MFR} with the spectra of social avalanches of knowledge creation, Fig.\ \ref{fig-Fluct-1}. In this case, the avalanche propagation is conditioned by the actor's activity patterns and the (non-)possession of the required expertise. According to Fig.\
\ref{fig-KCP-parameters}, a large number of users (agents) stops to be active after a given number of actions, while their artefacts are still available and can be the subject of interest to others. Thus, each avalanche becomes pinned by reaching the network node of such a user,  in the manner that a site defect pinns the moving domain wall.
Also, the users whose delay time is typically larger than the avalanche duration may have a similar effect on the current avalanche propagation. On the other hand, the
extremely active users (a small fraction, represented by the end of the distribution in Fig.\ \ref{fig-KCP-parameters}), accelerate the propagation through their numerous connections and also by being active at more than one question within a short time interval. This type of the actor's heterogeneity results in a
typical mixture of small and large events, as seen in the analysis in sec.\ \ref{sec-avalanches}. Beyond the shape of the multifractal spectra, representing a combination of small and large events,  the avalanche distributions differ in the random-site ferromagnetic model. Among other reasons, the evolving bipartite networks of the actors and their artefacts are identified as those being of the key importance.

\section{Discussion and Conclusions\label{sec-discussion}}
Considering the large dataset from Q\&A site Mathematics and the agent-directed modelling, we have analysed the avalanching behaviour and the bipartite network that underlies the creation of collective knowledge. Given the complexity of modelling the human actors, we have kept the agent's properties statistically similar to the features
of the users, detectable from the same empirical data. In particular, the agent's activity pattern is designed by the distributions of the number of actions per user $P(N_i)$ and the interactivity time $P(\Delta t)$ as well as the arrival rate $p(t)$, which are mutually  interconnected and characterise the human dynamics in the considered empirical system.
 We have varied the agent's expertise, as the most relevant feature to the knowledge creation. 
To evaluate its impact, the expertise matching to the contents of the artefact has been clearly observed in the simulations. (Other potential extensions of the model, e.g., neglecting or altering the above distributions independently, that goes beyond human dynamics \cite{BT-ABMbook}, are not
discussed in this work.)

The analysis indicates that the knowledge-creation process represents a particular class of social dynamics, in which the self-tuning towards the criticality is controlled by the use of knowledge in the \textit{meaningful} actions and the co-evolution of the underlying network.  Our key findings are here discussed.

\textit{The self-organised criticality.} The robustly observed temporal correlations,
avalanching and multifractality, as well as the scale-invariance dependence on the driving rate, indicate that the criticality might occur in these stochastic processes in a
self-tunned manner. The social dynamics, driven by the arrival of new actors and innovation that they bring, represents the
main source of the avalanching behaviours. Whereas, the altered degree of branching (e.g., in $\mu$-process) and, more importantly,  the strict use of the expertise imposes the constraints to the social dynamics, which
affects the avalanche propagation. These constraints then manifest in the non-universal scaling exponents of the avalanche sizes and durations. Moreover, the
relative fraction of the small and large avalanches, which appear to be mixed in the course of the process,  depends on
the available expertise of the actors, thus affecting the width of the multifractal spectrum. Non-extensivity ($q_a>1$) is another remarkable feature of the knowledge-creation process, where the $q_a$-exponential distribution applies to the avalanche sizes and the $q_r$-Gaussian distribution to the avalanche returns. Although the observed multifractality of the avalanche series
is compatible with these distributions, more theoretical work is needed to unravel their origin.

\textit{The structure of network partitions.} In this process, the growth of the bipartite network occurs by the addition of layers, which contain new arrivals and other active agents,  and their artefacts. 
How the network will grow is strongly related to the expertise-conditioned linking. After a sufficient time, the network exhibits a  broad distribution of the degree of nodes in both partitions.  The scaling exponent of the agent's degree distribution in each case is close to the introduced distribution of the activity  $P(N_i)$. While, the cut-offs of the distributions, indicating the actual size of the network, vary with the considered apportionment of the expertise. Notably, the network size increases when the
agents possess a larger expertise in the average. In this case, the probability of an agent to connect to a suitable artefact is elevated. Similarly, the network growth is accelerated by the addition of a larger number of actors in the initial stages of the process, which results in a greater number of the available artefacts. 
Thus, in the social process, the scale-freeness of the user partition is determined by the actual activity profiles alone. Where the corresponding edges will appear in the network, and, consequently the network's community structure \cite{we_SciRep2015},  depends on the available expertise and how it is distributed over the actors.  
On the other hand, the question-partition degree distribution strongly depends on the expertise. Note that in both cases the determined scaling exponents are smaller than 2. These results suggest that the effective mechanisms might be different from the popular 'preferential' attachment-and-rewiring rule that leads to the exponents larger than 2.

\textit{The self-tuning factors.} The heterogeneity of the actor's profiles and the activity patterns, with the extended range of delay times and the number of actions, is certainly an important factor for the appearance of the avalanches in the studied social dynamics. Apart from these 'human factors', it is stipulated that some other ingredients of the knowledge creation may contribute to the
paths towards the criticality. 
Specifically,   the importance and the use of expertise in the human collaborative endeavour makes the social process substantially different from, for instance, the activity on popular Blogs, where the negative emotional charge of comments may lead to a supercritical avalanche \cite{mitrovic2010a,we-entropy}. Here, the required expertise of the actors needs to match the contents of the question, resulting in a balanced activity (resembling the energy balance in the driven physical system). Consequently, the activity stops when sufficient knowledge is built through the answers on a particular question, depending on the available expertise of the actors. The same artefacts may become a focus in the later stages when new necessary knowledge becomes available, i.e., by the arrival of new players. 
Thus, in contrast to the standard social dynamics, a kind of
optimisation of the available expertise applies, which is
also compatible with the non-extensive dynamics mentioned above.
Note that the optimisation of the system's efficiency is often associated with the functioning of  biological systems \cite{SOC-biology2011} and the avalanching process in neuronal assemblies \cite{SOC-neuronalNat2007,SOC_neuroCulture2013} and the brain \cite{SOC-Brain-review2014}, which are still not well understood.
Furthermore, the underlying network evolution by the addition of the innovative contents in the active layer can be seen as another decisive factor to provide a particular type of critical behaviour. 
Theoretically, changing the random environment for the self-organising process affects the universality of the critical behaviour that can be achieved. The renormalization group study of the critical sandpile model in the presence of quenched \cite{BT_disorderinducedcrit1998} or annealed \cite{SOC-RG-Antonov2016} random currents have demonstrated that a new stable fixed point appears, which is controlled by the variance of the random variable. 
 For the knowledge-creation, it is relevant to mention that the innovation expansion builds the network of contents with a logical structure, which originates in the participant's individual knowledge. 
For the same empirical data, this aspect of knowledge creation
was demonstrated in \cite{we-KCNets_PLOS2016} by analysis of the knowledge network (containing the cognitive contents that are used in all questions and answers, and encoded by the standard mathematical classification scheme). 

Our study opens some questions for further theoretical considerations, in particular: the formal differences between knowledge-creation and common social dynamics; the potential similarity between the knowledge-creation and the brain avalanching dynamics; the origin of the non-extensivity (although  the non-extensive character of the dynamics is intuitive in the context of knowledge, the formal origin of the $q$-Gaussian fluctuations is not understood), and other issues. 
We hope that this work will initiate more intensive research towards deeper understanding (and possible control) of the avalanching processes in social systems. The presented results, based on the empirical data and the agent-directed model, which is almost equally complex as the empirical  system itself, reveal many factors that act in unison and contribute to the observed SOC. The presented comparison to the driven spin system with site defects suggests that apart from the expertise, the heterogeneity of the actor's activity patterns is an essential factor that prevents the appearance of the supercritical avalanches. 
Our findings may help design formal theoretical models of SOC, e.g., of the cellular automata type or the continuous models suitable for the renormalisation group analysis,  which may be capable of describing the unique role of each of these factors.

\section*{\normalsize Acknowledgments}
The authors acknowledge the financial support from the Slovenian
Research Agency (research code funding number P1-0044) and in part by the Projects 
ON17017 by the Ministry
of Education, Science and Technological Development of the Republic of Serbia
MMD and RM  also thank for kind hospitality during the stay at the
Department of Theoretical Physics, Jo\v zef Stefan Institute, where
this work was done.


\end{document}